\documentclass[letterpaper, 10 pt, conference]{ieeeconf}
    
    \IEEEoverridecommandlockouts 
    \makeatletter

    \let\proof\@undefined
    \let\endproof\@undefined
    \makeatother
    \let\labelindent\relax

    \usepackage{jhoagg}
    \usepackage{mathtools}
    \usepackage{amsfonts}
    \usepackage{amssymb,latexsym}
    \usepackage[psamsfonts]{eucal}
    \usepackage{amsthm}
    \usepackage{graphicx}
    \usepackage[inline]{enumitem}
    \usepackage{indentfirst}
    \usepackage{setspace}
    \usepackage{microtype}
    \usepackage{threeparttable}
    \usepackage{float}
    \usepackage{cleveref}
    \usepackage{afterpage}
    \usepackage{placeins}
    \usepackage{cancel}
    \usepackage{cite}
    \usepackage{leftidx}
    \usepackage{breqn}
    \usepackage[export]{adjustbox}
    \usepackage{comment}
    \usepackage[ruled, linesnumbered, lined]{algorithm2e}
    \usepackage[dvipsnames]{xcolor}
    \usepackage{xcolor}
    \usepackage{soul}

    \usepackage{tikz}
    \usepackage{pgf}
    \usepackage{pgfplots}
    \usepgflibrary{plothandlers}
    \usepgfplotslibrary{external} 
    \pgfkeys{/pgf/number format/.cd,1000 sep={}}  
    \usetikzlibrary{calc,shapes.misc,plotmarks}
    \pgfplotsset{compat=1.13}
    
    \let\originalleft\left
    \let\originalright\right
    \renewcommand{\left}{\mathopen{}\mathclose\bgroup\originalleft}
    \renewcommand{\right}{\aftergroup\egroup\originalright}

    \newcounter{thm} 
    \newtheorem{theorem}[thm]{\indent Theorem}
    
    \newtheorem{assumption}{\indent Assumption}
    
    \newtheorem{proposition}{\indent Proposition}
    
    \newtheorem{lemma}{\indent Lemma}
    
    \newtheorem{corollary}{\indent Corollary}
    
    \newtheorem{definition}{\indent Definition}
    
    \newtheorem{example}{\indent Example}
    
    \newtheorem{fact}{\indent Fact}
    
    \newtheorem{conjecture}{\indent Conjecture}
    
    \newtheorem{experiment}{\indent Experiment}

    \allowdisplaybreaks

    \renewcommand{\theenumi}{{\it (\alph{enumi})}}
    \renewcommand{\labelenumi}{\theenumi}

    \usepackage{cite}
    \usepackage{accents}
    
    \newlength\figureheight 
    \newlength\figurewidth

    \allowdisplaybreaks
    
    \graphicspath{ {Figures/} }
    \DeclareGraphicsExtensions{.png}
    
    \DeclareMathAlphabet{\mathcal}{OMS}{cmsy}{m}{n} 

    \crefname{equation}{}{}
    
    \usepackage[ruled, linesnumbered, lined]{algorithm2e}
    \SetKwInput{KwInit}{Initialize}
    \SetKwFunction{SafeSet}{SafeSet}
    \SetKwFunction{GetSafe}{GetSafe}
    \SetKwFunction{GetUnsafe}{GetUnsafe}
    
    \SetCommentSty{mycommfont}
    \newcommand{\ubar}[1]{\underaccent{\bar}{#1}}

\usepackage{fancyhdr}

\fancypagestyle{firststyle}
{
    \fancyfoot{} 
   \fancyfoot[L]{\textit{American Control Conference (ACC) 2023, May 31 - June 2, 2023.}}
}
    
\begin{document}
\title{Soft-Minimum Barrier Functions for Safety-Critical Control Subject to Actuation Constraints 
}
\author{Pedram Rabiee and Jesse B. Hoagg
\thanks{P. Rabiee and J. B. Hoagg are with the Department of Mechanical and Aerospace Engineering, University of Kentucky, Lexington, KY, USA. (e-mail: pedram.rabiee@uky.edu, jesse.hoagg@uky.edu).}
\thanks{This work is supported in part by the National Science Foundation (1849213) and Air Force Office of Scientific Research (FA9550-20-1-0028).}
}
    
\maketitle
\thispagestyle{firststyle}

\begin{abstract}
This paper presents a new control approach for guaranteed safety (remaining in a safe set) subject to actuator constraints (the control is in a convex polytope). 
The control signals are computed using real-time optimization, including linear and quadratic programs subject to affine constraints, which are shown to be feasible.  
The control method relies on a new soft-minimum barrier function that is constructed using a finite-time-horizon prediction of the system trajectories under a known backup control. 
The main result shows that: (i) the control is continuous and satisfies the actuator constraints, and (ii) a subset of the safe set is forward invariant under the control.
We also demonstrate this control on numerical simulations of an inverted pendulum and a double-integrator ground robot.
\end{abstract}


\section{Introduction}
\label{sec:Introduction}

Robots and autonomous systems are often required to respect safety-critical constraints while simultaneously achieving a specified task \cite{borrmann2015control,nguyen2015safety}. 
Safety constraints can be achieved by determining a control that makes a designated safe set $\SSS_\rms$ forward invariant with respect to the closed-loop dynamics~\cite{blanchini1999set}, that is, designing a control for which the state is guaranteed to remain inside $\SSS_\rms$.
%
%
%
Approaches that address safety using set invariance include reachability methods~\cite{chen2018hamilton}, model predictive control~\cite{wabersich2018safe}, and barrier function (BF) methods (e.g.,~\cite{prajna2007framework,panagou2015distributed,tee2009barrier,jin2018adaptive,ames2014control,ames2016control,ames2019control}). 

In particular, BFs have been employed in a variety of ways. 
For example, they have been used for Lyapunov-like control design and analysis \cite{prajna2007framework,panagou2015distributed,tee2009barrier,jin2018adaptive}.
In contrast, the control barrier function (CBF) approaches in~\cite{ames2014control,ames2016control,ames2019control} compute the control signal using real-time optimization.
These optimization-based methods are modular in that 
they combine a nominal performance controller (which may not attempt to respect safety) with a safety filter that performs a real-time optimization using CBF-based constraints to generate a control that guarantees safety.  
This real-time optimization is often formulated as an instantaneous minimum-intervention problem, that is, the problem of finding a control at the current time that is as close as possible to the nominal performance control while satisfying the CBF-based safety constraints.

Barrier-function methods typically rely on the assumption that $\SSS_\rms$ is control forward invariant (i.e., there exists a control that makes $\SSS_\rms$ forward invariant). 
For systems without actuator constraints (i.e., input constraints), control forward invariance is satisfied under relatively minor structural assumptions (e.g., constant relative degree). 
In this case, the control can be generated from a quadratic program that employs feasible CBF-based constraints (e.g., \cite{ames2014control,ames2016control,ames2019control}).
In contrast, actuator constraints can prevent $\SSS_\rms$ from being control forward invariant.
In this case, it may be possible to compute a control forward invariant subset of $\SSS_\rms$ using methods such as Minkowski operations~\cite{rakovic2004computation}, sum-of-squares~\cite{korda2014convex, xu2017correctness}, approximate solutions of a Hamilton-Jacobi partial differential equation~\cite{mitchell2005time}, or sampling~\cite{gillula2014sampling}. 
However, these methods may not scale to high-dimensional systems.

Another approach to address safety with actuator constraints is to use a prediction of the system trajectories into the future to obtain a control forward invariant subset of $\SSS_\rms$. 
For example,~\cite{squires2018constructive} uses the trajectory under a backup control to address safety with actuator constraints. 
However,~\cite{squires2018constructive} uses an infinite time horizon prediction. 
In contrast,~\cite{gurrietScalableSafety2020,chen2020guaranteed} determine a control forward invariant subset of $\SSS_\rms$ from a BF constructed from a finite-horizon prediction under a backup control.
This BF uses the minimum function; thus, it is not continuously differentiable and cannot be used directly to form a BF-based constraint for real-time optimization.
Instead,~\cite{gurrietScalableSafety2020,chen2020guaranteed} replace the original BF by a finite number of continuously differentiable BFs---each of which are used to form BF-based constraints for real-time optimization. 
However, the number of substitute BFs (and thus optimization constraints) increases as the prediction horizon increases. 
In addition, the multiple BF-based constraints can be conservative.
Finally, it is worth noting that~\cite{gurrietScalableSafety2020,chen2020guaranteed} do not guarantee feasibility of these multiple BF-based constraints.

This paper presents a novel soft-minimum BF that uses a finite-horizon prediction of the system's trajectory under a backup control.
We show that this BF describes a control forward invariant (subject to actuator constraints) subset of $\SSS_\rms$.
Since the soft-minimum BF is continuously differentiable, it can be used to form a single non-conservative BF-based constraint for optimization regardless of the prediction horizon. 
The advantages of the soft-minimum BF facilitate the paper's second contribution, namely, a real-time optimization-based control that guarantees safety with actuator constraints.
Notably, the control is continuous, and the required optimization is convex with feasible constraints.

\section{Soft Minimum}

Let $\rho>0$, and consider the function $\mbox{softmin}_\rho : \BBR \times \cdots \times \BBR \to \BBR$ defined by 
\begin{equation}
\mbox{softmin}_\rho (z_1,\ldots,z_N) \triangleq -\frac{1}{\rho}\log\sum_{i=1}^Ne^{-\rho z_i},
\end{equation}
which is the \textit{soft minimum}. 
The next result shows that soft minimum is a lower bound on minimum.

\begin{fact} \label{fact:softmin_bound}
\rm{
Let $z_1,\ldots, z_N \in \BBR$. 
Then,
\begin{align*}
 \min \, \{ z_1,\ldots, z_N \} - \frac{\log N }{\rho} 
 &\le \mbox{softmin}_\rho(z_1,\ldots,z_N) \\
 &< \min \, \{z_1,\ldots, z_N\},
\end{align*}
}
\end{fact}

Fact~\ref{fact:softmin_bound} shows that as $\rho \to \infty$, $\mbox{softmin}_\rho$ converges to the minimum. 
Thus, $\mbox{softmin}_\rho$ is a smooth approximation of the minimum.




\section{Problem Formulation}

Consider the system
\begin{equation}\label{eq:dynamics}
\dot x(t) = f(x(t))+g(x(t)) u(t),
\end{equation}
where $x(t) \in \BBR^{n}$ is the state, $x_0 = x(0)$ is the initial condition, and $u(t) \in \BBR^m$ is the control.
Let $A_u \in \BBR^{r \times m}$ and $b_u \in \BBR^{r}$, and define
\begin{equation}
    \SU \triangleq \{u \in \BBR^m: A_u u \le b_u\} \subset \BBR^{m},
    \label{eq:SU}
\end{equation}
which is the set of admissible controls. 
We assume that $\SU$ is bounded and not empty.
We call $u$ an \textit{admissible control} if for all $t \ge 0$, $u(t) \in \SU$.

Let $h_\rms:\BBR^n \to \BBR$ be continuously differentiable, and define the \textit{safe set}
\begin{equation}
\SSS_\rms \triangleq \{x \in \BBR^n \colon h_\rms(x) \geq 0 \}.
\label{eq:S_s}
\end{equation}
Note that $\SSS_\rms$ is not assumed to be control forward invariant with respect to \eqref{eq:dynamics} where $u$ is an admissible control.
In other words, there may not exist an admissible control $u$ such that if $x_0 \in \SSS_\rms$, then for all $t \ge 0$, $x(t) \in \SSS_\rms$.

Next, consider the \textit{desired control} $u_\rmd : [0,\infty) \to \BBR^m$.
We note that $u_\rmd$ is not necessarily an admissible control. 
In addition, $\SSS_\rms$ is not necessarily forward invariant with respect to \eqref{eq:dynamics} where $u = u_\rmd$. 

The objective is to design a full-state feedback control $u:\BBR^n \to \BBR^m$ such that for all initial conditions in a subset of $\SSS_\rms$, the following hold:

\begin{enumerate}[leftmargin=0.9cm]
	\renewcommand{\labelenumi}{(O\arabic{enumi})}
	\renewcommand{\theenumi}{(O\arabic{enumi})}

 \item\label{obj1}
For all $t \ge 0$, $x(t) \in \SSS_\rms$.

\item\label{obj2}
For all $t \ge 0$, $u(x(t)) \in \SU$.

\item\label{obj3}
For all $t \ge 0$, $\| u(x(t)) -u_\rmd(t) \|_2$ is small.
\end{enumerate}

The following notation is needed. 
For a continuously differentiable function $\eta:\BBR^n \to \BBR$, the Lie derivatives of $\eta$ along the vector fields of $f$ and $g$ are defined as 
\begin{equation*}
L_f \eta(x) \triangleq \frac{\partial \eta(x)}{\partial x}f(x), \qquad L_g \eta(x) \triangleq \frac{\partial \eta(x)}{\partial x}g(x).
\end{equation*}

\section{Preliminary Results on Barrier Functions Using Trajectory under Backup Control}\label{section:softmin_prelim}

Consider a continuously differentiable \textit{backup control} $u_\rmb : \BBR^n \to \SU$.
Let $h_\rmb : \BBR^n \to \BBR$ be continuously differentiable, and 
define the \textit{backup safe set}
\begin{equation}
\SSS_\rmb \triangleq \{x \in \BBR^n \colon h_\rmb(x) \geq 0 \}.
\end{equation}
We assume $\SSS_\rmb \subseteq \SSS_\rms$ and make the following assumption.

\begin{assumption}\label{assump:S_b_contract}\rm
If $u=u_\rmb$ and $x_0 \in \SSS_\rmb$, then for all $t \ge 0$, $x(t) \in \SSS_\rmb$. 
\end{assumption}


Assumption~\ref{assump:S_b_contract} states that $\SSS_\rmb$ is forward invariant with respect to \eqref{eq:dynamics} where $u = u_\rmb$.
However, $\SSS_\rmb$ may be small relative to $\SSS_\rms$.

Consider $\tilde f: \BBR^n \to \BBR^n$ defined by 
\begin{equation}\label{eq:ftilde}
\tilde f(x) \triangleq f(x) + g(x) u_\rmb(x),
\end{equation}
which is the right-hand side of the closed-loop dynamics under the backup control $u_\rmb$. 
Next, let $\phi: \BBR^n \times [0,\infty) \to \BBR^n$ satisfy 
\begin{equation}\label{eq:phi_def}
    \phi(x, \tau) = x +\int_{0}^{\tau} \tilde f(\phi(x,\sigma)) \, \rmd \sigma,
\end{equation}
which implies that $\phi(x,\tau)$ is the solution to \eqref{eq:dynamics} at time $\tau$ with $u=u_\rmb$ and initial condition $x$.

Let $T > 0$ be a time horizon, and consider $h_*:\BBR^n \to \BBR$ defined by
\begin{equation}
h_*(x) \triangleq \min \, \left \{ h_\rmb(\phi(x,T)),  \min_{\tau \in [0,T]} h_\rms(\phi(x,\tau)) \right \}, \label{eq:h_min_cont_def}
\end{equation}
and define
\begin{equation}
\SSS_* \triangleq \{ x \in \BBR^n \colon h_*(x) \ge 0 \}\label{eq:defS_cont}.
\end{equation}
Note that for all $x \in \SSS_*$, the solution \eqref{eq:phi_def} under $u_\rmb$ does not leave $\SSS_\rms$ and reaches $\SSS_\rmb$ within time $T$. 
The next result relates $\SSS_*$ to $\SSS_\rmb$ and $\SSS_\rms$.
The result is similar to \cite[Proposition 6]{gurrietScalableSafety2020}. 
The proof is omitted due to space limitations.

\begin{proposition} \label{prop:S*}
\rm $\SSS_\rmb \subseteq \SSS_* \subseteq \SSS_\rms$.
\end{proposition}



The next result shows that $\SSS_*$ is forward invariant with respect to \eqref{eq:dynamics} where $u=u_\rmb$. 
In fact, not only is $\SSS_*$ forward invariant but the state converges to $\SSS_\rmb \subseteq \SSS_*$ by time $T$. 
The proof is omitted due to space limitations.

\begin{proposition} \label{prop:fwd_invar_Sc}
\rm 
Consider \eqref{eq:dynamics}, where $u = u_\rmb$ and $x_0 \in \SSS_*$.
Then, the following statements hold:
\begin{enumerate}
\item For all $t \ge T$, $x(t) \in \SSS_\rmb$. \label{fact:fwd_invar_Sc_Sb}
\item For all $t \ge 0$, $x(t) \in \SSS_*$. \label{fact:fwd_invar_Sc_traj}
\end{enumerate}
\end{proposition}




Since Proposition~\ref{prop:fwd_invar_Sc} implies that $\SSS_*$ is forward invariant with $u=u_\rmb$ and $u_\rmb$ is admissible, it follows that for all $x_0 \in \SSS_*$, the backup control $u_\rmb$ satisfies \ref{obj1} and \ref{obj2}. 
However, $u_\rmb$ does not address \ref{obj3}.
One approach to address \ref{obj3} is to use $h_*$ as a BF in a minimum intervention quadratic program. 
However, $h_*$ is not continuously differentiable and cannot be used directly to construct a BF-based constraint. 
Instead of using $h_*$ as the BF, \cite{gurrietScalableSafety2020} uses multiple BFs---one for each argument of the minimum in \eqref{eq:h_min_cont_def}.
However, \eqref{eq:h_min_cont_def} has an infinite number of arguments because the minimum is over $[0,T]$.
This issue is addressed in \cite{gurrietScalableSafety2020} by using a sampling of times. 
Specifically, let $N$ be a positive integer, and let $T_\rms \triangleq T/N$. 
Then, consider $\bar h_*:\BBR^n \to \BBR$ defined by
\begin{equation}
\bar h_*(x) \triangleq \min \, \left \{ h_\rmb(\phi(x,NT_\rms)),  \min_{i \in \{ 0,1,\ldots,N\} } h_\rms(\phi(x,iT_\rms)) \right \},
\label{eq:h_min_def}
\end{equation}
and define
\begin{equation}
\bar \SSS_* \triangleq \{ x \in \BBR^n \colon \bar h_*(x) \ge 0 \}. \label{eq:defS_star}
\end{equation}

The next result relates $\bar \SSS_*$ to $\SSS_*$ and $\SSS_\rms$.
The proof is omitted due to space limitations.

\begin{proposition} \label{prop:bar S*}
\rm $\SSS_* \subseteq \bar \SSS_* \subseteq \SSS_\rms$.
\end{proposition}



The next result shows that for all $x_0 \in \bar \SSS_*$, the backup control $u_\rmb$ forces $x$ to converge to $\SSS_\rmb$ by time $T$.
In addition, the result shows that for all $x_0 \in \bar \SSS_*$, the state is in $\bar \SSS_*$ at the sample times $T_\rms,2T_\rms,\ldots,N T_\rms$.
The proof is omitted due to space limitations.

\begin{proposition} \label{prop:fwd_invar_S*}
\rm 
Consider \eqref{eq:dynamics}, where $u = u_\rmb$ and $x_0 \in \bar \SSS_*$. 
Then, the following statements hold:
\begin{enumerate}
\item For all $t \ge T$, $x(t) \in \SSS_\rmb$. \label{fact:fwd_invar_S*_Sb}
\item For all $i\in \{ 0,1,\ldots,N \}$, $x(iT_\rms) \in \bar \SSS_*$. \label{fact:fwd_invar_S*_traj}
\end{enumerate}
\end{proposition}

Proposition~\ref{prop:fwd_invar_S*} does not provide any information about the state in between the sample times. 
Thus, Proposition~\ref{prop:fwd_invar_S*} does not imply that $\bar \SSS_*$ is forward invariant with respect to \eqref{eq:dynamics} with $u=u_\rmb$. 
However, we can determine a superlevel set of $\bar h_*$ such that for all initial conditions in that superlevel set, $u_\rmb$ keeps the state in $\SSS_*$ for all time. 
To define this superlevel set, let $l_\rms$ be the Lipschitz constant of $h_\rms$ with respect to the Euclidean norm, and define
\begin{equation*}
l_\phi \triangleq \sup_{x\in \bar \SSS_*} \| f(x) + g(x)u_\rmb(x)\|_2,    
\end{equation*}
which is finite if $\SSS_\rms$ is bounded. 
Define the superlevel set
\begin{equation}\label{eq:ubarS_*_def}
\ubar \SSS_* \triangleq \left \{x\in \BBR^n : \bar h_*(x) \ge \tfrac{1}{2} T_\rms l_\phi l_\rms \right \}.
\end{equation}
The following result shows that $\ubar \SSS_*$ is a subset of $\SSS_*$. 
The proof is omitted due to space limitations; however, it relies, in part, on arguments similar to those in \cite[Theorem 1]{gurrietScalableSafety2020}.

\begin{proposition}\label{prop:set_rel_softmin}
\rm 
$\ubar \SSS_* \subseteq \SSS_* \subseteq \bar \SSS_* \subseteq\SSS_\rms$. 
\end{proposition}

Together, Propositions~\ref{prop:fwd_invar_Sc} and \ref{prop:set_rel_softmin} imply that for all $x_0 \in \ubar \SSS_*$, the backup control $u_\rmb$ keeps the state in $\SSS_*$ for all time. 
Thus, $u_\rmb$ satisfies \ref{obj1} and \ref{obj2} but does not address \ref{obj3}.

Since $\bar h_*$ is not continuously differentiable, \cite{gurrietScalableSafety2020} addresses \ref{obj3} using a minimum intervention quadratic program with $N+1$ BFs---one for each of the arguments in \eqref{eq:h_min_def}. 
However, this approach has 3 drawbacks. 
First, the number of BFs increases as the time horizon $T$ increases or the sample time $T_\rms$ decreases (i.e., as $N$ increases). 
Thus, the number of affine constraints and computational complexity increases as $N$ increases. 
Second, although imposing an affine constraint for each of the $N+1$ BFs is sufficient to ensure that $\bar h_*$ remains positive, it is not necessary. 
In particular, these $N+1$ affine constraints are conservative. 
In other words, the $N+1$ affine constraints can significantly limit the set of feasible control. 
Third, the method in \cite{gurrietScalableSafety2020} does not guarantee feasibility of the quadratic program solved to obtain the control. 
In the next subsection, we use a soft minimum BF to approximate $\bar h_*$ and present a control synthesis approach with guaranteed feasibility and where the number of affine constraints is fixed (i.e., independent of $N$).

\section{Safety-Critical Control Using Soft Minimum Barrier Function}\label{section:softmin}

This section presents a continuous control that guarantees safety subject to the constraint that the control is admissible (i.e., contained in $\SU$). 
The control is computed using a minimum intervention quadratic program with a soft minimum BF constraint. 
The control also relies on a linear program to provide a feasibility metric, that is, a measure of how close the quadratic program is to becoming infeasible. 
Then, the control continuously transitions to the backup control $u_\rmb$ if the feasibility metric or the soft minimum BF is less than a user-defined threshold.

Let $\rho > 0$, and consider $h:\BBR^n \to \BBR$ defined by
\begin{align}
h(x) &\triangleq  \mbox{softmin}_{\rho} ( h_\rms(\phi(x,0)), h_\rms(\phi(x,T_\rms)),h_\rms(\phi(x,2 T_\rms)), \nn\\ 
&\qquad \ldots, h_\rms(\phi(x,N T_\rms)), h_\rmb(\phi(x,NT_\rms))),\label{eq:h_softmin_def}
\end{align}
which is continuously differentiable. 
Define
\begin{equation}\label{eq:defS_softmin}
\SSS \triangleq \{ x \in \BBR^n \colon h(x) \ge 0 \}.
\end{equation}
Fact~\ref{fact:softmin_bound} implies that for all $x \in \BBR^n$, $h(x) < \bar h_*(x)$. 
Thus, $\SSS \subset \bar \SSS_*$.
Fact~\ref{fact:softmin_bound} also implies that for sufficiently large $\rho>0$, $h(x)$ is arbitrarily close to $\bar h_*(x)$. 
Thus, $h$ is a smooth approximation of $\bar h_*$.
Note that the if $\rho >0$ is small, then $h$ can be a conservative approximation of $\bar h_*$.
In contrast, if $\rho >0$ is large, then $h$ is a less conservative approximation of $\bar h_*$.
However, in this case, $\frac{\partial h(x)}{\partial x}$ can have a large norm at the points where $\bar h_*$ is not differentiable.
Thus, selecting $\rho$ is a trade-off between the conservativeness of $h$ and the magnitude of the norm of $\frac{\partial h(x)}{\partial x}$.



Next, let $\alpha >0$ and $\epsilon \in [0, \sup_{x\in\SSS} h(x))$. 
Consider  $\beta\colon \BBR^n \to \BBR$ defined by
\begin{equation}
\label{eq:feas_check}
\beta(x) \triangleq  L_f h(x) + \alpha (h(x)- \epsilon) + \max_{\hat u \in \SU} 
 L_g h(x)  \hat u,
\end{equation}
where for all $x \in \BBR^n$, $\beta(x)$ exists because $\SU$ is not empty.
Define
\begin{equation}
\label{eq:feasible_set_def}
    \SB \triangleq \{ x \in\BBR^n \colon \beta(x) \ge 0 \},
\end{equation}
and the next result follows immediately from \cref{eq:feas_check,eq:feasible_set_def}.

\begin{proposition}\label{prop:feas_softmin}
{\rm
For all $x \in \SB$, there exists $\hat u \in \SU$ such that
\begin{equation*}
L_f h(x) + L_g h(x) \hat u + \alpha (h(x) - \epsilon) \ge 0.
\end{equation*}
}
\end{proposition}


Consider $\gamma \colon \BBR^n \to \BBR$ defined by
\begin{equation}\label{eq:gamma_def}
\gamma(x) \triangleq \min\{h(x) - \epsilon, \beta(x)\},
\end{equation}
and define
\begin{equation}\label{eq:Gamma_def}
    \Gamma \triangleq \{ x \in\BBR^n \colon \gamma(x) \ge 0 \}.
\end{equation}
Note that $\Gamma \subseteq \SB$.
For all $x \in \Gamma$, define
\begin{subequations}\label{eq:qp_softmin}
\begin{align}
& u_*(x) \triangleq \underset{\hat u \in \SU}{\mbox{argmin}}  \, 
\|\hat u - u_\rmd(x)\|^2 \label{eq:qp_softmin.a}\\
& \text{subject to}\nn\\
& L_f h(x) + L_g h(x) \hat u + \alpha (h(x) - \epsilon) \ge 0. \label{eq:qp_softmin.b}
\end{align}
\end{subequations}
Since $\Gamma \subseteq \SB$, Proposition~\ref{prop:feas_softmin} implies that for all $x \in \Gamma$, the quadratic program \eqref{eq:qp_softmin} has a solution.

Let $\kappa > 0$, and consider a continuous function $\sigma:\BBR \to [0,1]$ such that for all $a \in (-\infty,0]$, $\sigma(a) =0$; for all $a \in [\kappa,\infty)$, $\sigma(a) =1$; and $\sigma$ is strictly increasing on $a \in [0,\kappa]$. 
The following example provides one possible choice for $\sigma$. 

\begin{example}\label{ex:sigma}\rm
Consider $\sigma:\BBR \to [0,1]$ given by
\begin{equation*}
    \sigma(a) = \begin{cases}
    0, & \mbox{if } a < 0, \\
    \frac{a}{\kappa}, & \mbox{if } 0 \le a \le \kappa, \\
    1, & \mbox{if } a > \kappa. \\
    \end{cases}
\end{equation*}
\end{example}

Finally, define the control 
\begin{equation}\label{eq:u_cases_softmin}
    u(x) = \begin{cases} [1-\sigma(\gamma(x))] u_\rmb(x) + \sigma(\gamma(x)) u_*(x), & \mbox{if } x \in \Gamma, \\ u_\rmb(x), & \mbox{else}.
    \end{cases}
\end{equation}

Since the soft-minimum BF $h$ is continuously differentiable, the quadratic program \eqref{eq:qp_softmin} requires only the single affine constraint \eqref{eq:qp_softmin.b} as opposed to the $N+1$ affine constraints used in \cite{gurrietScalableSafety2020}. 
Since \eqref{eq:qp_softmin} has only one affine constraint \eqref{eq:qp_softmin.b}, we can define the feasible set $\SB$ as the $0$-superlevel set of $\beta$, which relies on the solution to the linear program in \eqref{eq:feas_check}.  
Thus, since there is only one affine constraint, we can use the homotopy in \eqref{eq:u_cases_softmin} to smoothly transition from $u_*$ to $u_\rmb$ as $x$ leaves $\Gamma$.

The next theorem is the main result on the control~\Cref{eq:h_softmin_def,eq:defS_softmin,eq:feas_check,eq:feasible_set_def,eq:gamma_def,eq:Gamma_def,eq:qp_softmin,eq:u_cases_softmin} that uses the soft-minimum BF approach.
Note that $\mbox{bd\,}\SA$ denotes the boundary of the set $\SA$.

\begin{theorem}
\label{thm:softmin}
\rm 
Consider \eqref{eq:dynamics} and the control $u$ given by~\Cref{eq:h_softmin_def,eq:defS_softmin,eq:feas_check,eq:feasible_set_def,eq:gamma_def,eq:Gamma_def,eq:qp_softmin,eq:u_cases_softmin}, where $\SU$ is bounded, nonempty, and given by \eqref{eq:SU}, and $u_\rmb$ is continuously differentiable and satisfies Assumption~\ref{assump:S_b_contract}.
Then, the following conditions hold:
\begin{enumerate}

\item 
$u$ is continuous on $\BBR^n$.
\label{thm:softmin_u_continuity}

\item 
For all $x \in \BBR^n$, $u(x) \in \SU$. 
\label{thm:softmin_u_cond}

\item 
Assume $x(t) \in \rm{bd \,}\bar \SSS_*$. 
Then, there exists $\tau \in (0, T_\rms]$ such that $x(t+\tau) \in \bar \SSS_* \subseteq \SSS_\rms$.
\label{thm:softmin_return_to_S_star}

\item Assume $x_0 \in \SSS_*$, and let $\epsilon \ge \frac{1}{2} T_\rms l_\phi l_\rms$.
Then, for all $t \ge 0$, $x(t) \in \SSS_* \subseteq \SSS_\rms$. \label{thm:softmin_forward_inv}

\end{enumerate}
\end{theorem}

Parts~\ref{thm:softmin_u_continuity} and \ref{thm:softmin_u_cond} guarantee that the control is continuous and admissible. 
Part~\ref{thm:softmin_return_to_S_star} does not guarantee that $x$ stays in the safe set $\SSS_\rms$; however,~\ref{thm:softmin_return_to_S_star} implies that if $x$ leaves $\SSS_\rms$ in between sample times, then it must return to $\SSS_\rms$ by the next sample time.
This is a result of the fact that $h$ is constructed from a sampling of time. 
Finally,~\ref{thm:softmin_forward_inv} states that if $\epsilon \ge \frac{1}{2} T_\rms l_\phi l_\rms$, then, $\SSS_*$ is forward invariant under the control~\Cref{eq:h_softmin_def,eq:defS_softmin,eq:feas_check,eq:feasible_set_def,eq:gamma_def,eq:Gamma_def,eq:qp_softmin,eq:u_cases_softmin}.
In this case, $x$ is in the safe set $\SSS_\rms$ for all time.

The control~\Cref{eq:h_softmin_def,eq:defS_softmin,eq:feas_check,eq:feasible_set_def,eq:gamma_def,eq:Gamma_def,eq:qp_softmin,eq:u_cases_softmin} relies on the Lie derivatives in \cref{eq:feas_check,eq:qp_softmin.b}.
To calculate $L_fh$ and $L_gh$, note that 
\begin{align}
\label{eq:softmin_grad_h}
\frac{\partial h(x)}{\partial x} &= \frac{1}{e^{-\rho h(x)}} \left(\vphantom{\sum_{i=0}^N}
e^{-\rho h_\rmb(\phi(x,NT_\rms)} h_\rmb^\prime(\phi(x,NT_\rms))Q(x, NT_\rms) \right . \nn\\
&\quad \left .  +\sum_{i=0}^N e^{-\rho h_\rms(\phi(x,iT_\rms)} h_\rms^\prime(\phi(x,iT_\rms))Q(x, iT_\rms) \right ),
\end{align}
where $h_\rmb^\prime,h_\rms^\prime \colon \BBR^n \times \to \BBR^{1 \times n}$ are defined by
\begin{equation}\label{eq:h_s_h_b_grad}
h_\rmb^\prime(x) \triangleq \frac{\partial h_\rmb(x)}{\partial x}, 
\qquad h_\rms^\prime(x) \triangleq \frac{\partial h_\rms(x)}{\partial x},
\end{equation}
and $Q : \BBR^n \times [0,\infty) \to \BBR^{n \times n}$ is defined by
\begin{equation}\label{eq:Q_def}
Q(x,\tau) \triangleq \frac{\partial \phi(x,\tau)}{\partial x}.
\end{equation}
Differentiating \eqref{eq:phi_def} with respect to $x$ yields
\begin{equation}\label{eq:Q_integral}
    Q(x,\tau) = I + \int_0^\tau \tilde f^\prime (\phi(x,s)) Q (x,s) \, \rmd s,
\end{equation}
where $\tilde f^\prime \colon \BBR^n \to \BBR^{n \times n}$ is defined by $\tilde f^\prime(x) \triangleq {\partial \tilde f(x)}/{\partial x}$. 
Next, differentiating \eqref{eq:Q_integral} with respect to $\tau$ yields 
\begin{equation}\label{eq:sensitivity_ode}
    \frac{\partial Q(x,\tau)}{\partial \tau} = \tilde f^\prime (\phi(x,\tau)) Q(x,\tau).
\end{equation}
Note that for all $x \in \BBR^n$, $Q(x, \tau)$ is the solution to \eqref{eq:sensitivity_ode}, where the initial condition is $Q(x, 0)= I$. 
Thus, for all $x \in \BBR^n$, $L_fh(x)$ and $L_gh(x)$ can be calculated from \Cref{eq:softmin_grad_h}, where $\phi(x,\tau)$ is the solution to \eqref{eq:dynamics} under $u_\rmb$ on the interval $\tau \in [0,T]$ with $\phi(x,0) = x$, and $Q(x, \tau)$ is the solution to \eqref{eq:sensitivity_ode} on the interval $\tau \in [0,T]$ with $Q(x, 0)= I$. 
In practice, these solutions can be computed numerically at each time instant where the control algorithm \Cref{eq:u_cases_softmin} is executed (i.e., the time instants where the control is updated). 
Algorithm~\ref{alg:softmin} summarizes the implementation of  \Cref{eq:u_cases_softmin}, where $\delta t > 0$ is the time increment for a zero-order-hold on the control.

\begin{algorithm}[h!]\label{alg:softmin}
\DontPrintSemicolon
\caption{Control using the soft-minimum BF quadratic program}
\KwIn{ $u_\rmd$, $u_\rmb$, $h_\rmb$, $h_\rms$, $N$, $T_\rms$, $\epsilon$, $\kappa$, $\sigma$, $\delta t$, $\rho$
}

\For{$k=0,1,2,\ldots$}{
    $x \gets x(k\delta t)$\;
    Solve~\eqref{eq:phi_def} numerically to obtain $\{\phi(x,iT_\rms)\}_{i=0}^N$\;
    Solve~\eqref{eq:sensitivity_ode} numerically to obtain $\{Q(x,iT_\rms)\}_{i=0}^N$\;
    Compute $L_fh(x)$ and $L_gh(x)$ using~\cref{eq:softmin_grad_h,eq:h_s_h_b_grad}\;
    $h \gets$ \eqref{eq:h_softmin_def}\;
    $\beta \gets$ solution to linear program~\eqref{eq:feas_check}\;
    $\gamma \gets \min \{h- \epsilon, \beta\}$\;
     
    \eIf{$\gamma < 0$}{
        $u \gets u_\rmb(x)$\;}
    {
    $u_* \gets$ solution to quadratic program~\eqref{eq:qp_softmin}\;
    $u \gets [1-\sigma(\gamma)] u_\rmb(x) + \sigma(\gamma) u_*$\;
    }
}
\end{algorithm}

\section{Numerical Examples}

\textbf{Inverted Pendulum.} 
Consider the inverted pendulum modeled by \eqref{eq:dynamics}, where 
\begin{equation*}
    f(x) = \begin{bmatrix}
    \dot \theta \\
    \sin \theta
    \end{bmatrix}, 
    \qquad
    g(x) = \begin{bmatrix}
    0 \\
    1
    \end{bmatrix}, 
    \qquad 
    x = \begin{bmatrix}
    \theta \\
    \dot \theta
    \end{bmatrix}, 
\end{equation*}
and $\theta$ is the angle from the inverted equilibrium.
Let $\bar u = 1.5$ and $\SU = \{ u \in \BBR \colon u \in [-\bar u, \bar u] \}$.
The safe set is $\SSS_\rms$ is given by \eqref{eq:S_s}, where $h_\rms(x) = \pi - \|x\|_p$, $\|\cdot\|_p$ is the $p$-norm, and $p = 100$. 
The backup control is $u_\rmb(x) = \mbox{csat }K x$, where $\mbox{csat} \colon \BBR \to \SU$ is a continuously differentiable approximation of the saturation function, and $K = [ \, -3 \quad -3 \,]$.
Let $h_\rmb(x) = c_\rmb - x^\rmT P_\rmb x$, where $c_\rmb = 0.07$ and $P_\rmb = \matls 1.25 &  0.25 \\ 0.25 & 0.25 \matrs$, and note that it can be confirmed using Lyapunov's direct method that Assumption~\ref{assump:S_b_contract} is satisfied. The desired control is $u_\rmd = 0$, which implies that the objective is to stay inside $\SSS_\rms$ using instantaneously minimum control effort.

We implement the soft-minimum BF control~\Cref{eq:h_softmin_def,eq:defS_softmin,eq:feas_check,eq:feasible_set_def,eq:gamma_def,eq:Gamma_def,eq:qp_softmin,eq:u_cases_softmin}, where $\rho = 100$, $\alpha =1$, and $\sigma$ is given by \Cref{ex:sigma} where $\kappa = 0.05$. 
We let $\delta t= 0.1$~s, $N = 50$, and $T_\rms = 0.1$~s, which implies that the time horizon is $T = 5$~s.

\begin{figure}[t!]
\center{\includegraphics[width=0.48\textwidth,clip=true,trim= 0in 0.33in 1.0in 1.0in] {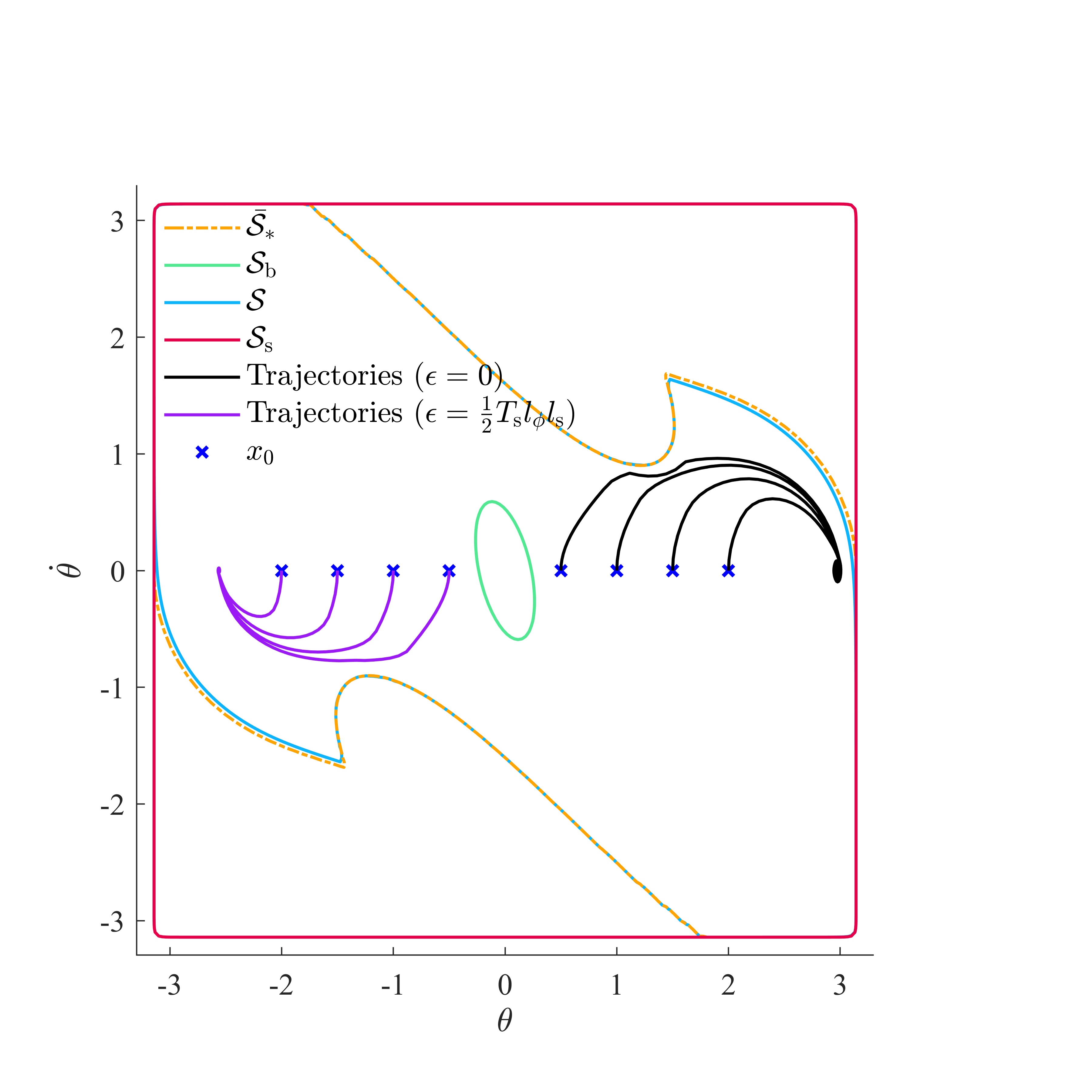}}
\caption{
$\SSS_\rms$, $\SSS_\rmb$, $\SSS$, $\bar \SSS_*$, and closed-loop trajectories for 8 initial conditions.}\label{fig:pendulum_traj}
\end{figure}

\Cref{fig:pendulum_traj} shows $\SSS_\rms$, $\SSS_\rmb$, $\SSS$, and $\bar \SSS_*$. 
Note that $\SSS \subset \bar \SSS_*$.
\Cref{fig:pendulum_traj} also provides the closed-loop trajectories for 8 initial conditions, specifically, $x_0 = [ \, \theta_0 \quad 0 \,]^\rmT$, where $\theta_0 \in \{ \pm 0.5, \pm 1, \pm 1.5, \pm 2\}$.
We let $\epsilon =0$ for the initial conditions with $\theta_0 \in \{  0.5, 1, 1.5, 2\}$, and we let $\epsilon = \frac{1}{2} T_\rms l_\phi l_\rms$ for the initial conditions with $\theta_0 \in \{  -0.5, -1, -1.5, -2\}$, which are the reflection of the first 4 across the origin.
For the cases with $\epsilon = \frac{1}{2} T_\rms l_\phi l_\rms$, part \ref{thm:softmin_forward_inv} of \Cref{thm:softmin} implies that $\SSS_*$ is forward invariant under the control~\Cref{eq:u_cases_softmin}.
Note that the trajectories with $\epsilon = \frac{1}{2} T_\rms l_\phi l_\rms$ are more conservative than those with $\epsilon = 0$.

\begin{figure}[t!]
\center{\includegraphics[width=0.48\textwidth,clip=true,trim= 0.25in 0.25in 1.1in 0.6in] {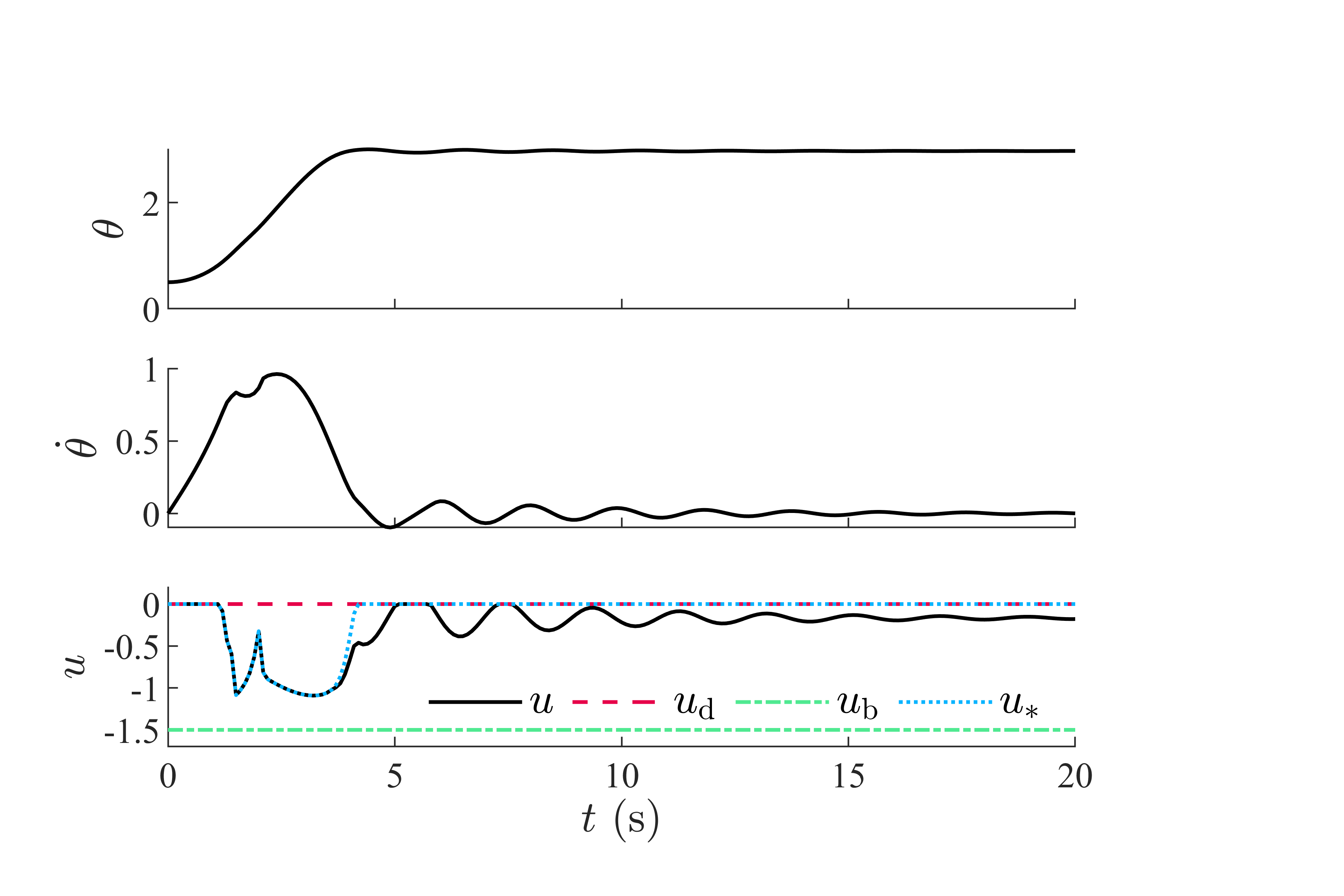}}
\caption{$\theta$, $\dot \theta$, $u$, $u_\rmd$, $u_\rmb$ and $u_*$ for $x_0=[0.5\,\,0]^\rmT$ and $\epsilon = 0$.}\label{fig:pendulum_states}
\end{figure}

\begin{figure}[t!]
\center{\includegraphics[width=0.48\textwidth,clip=true,trim= 0.25in 0.2in 1.1in 0.5in] {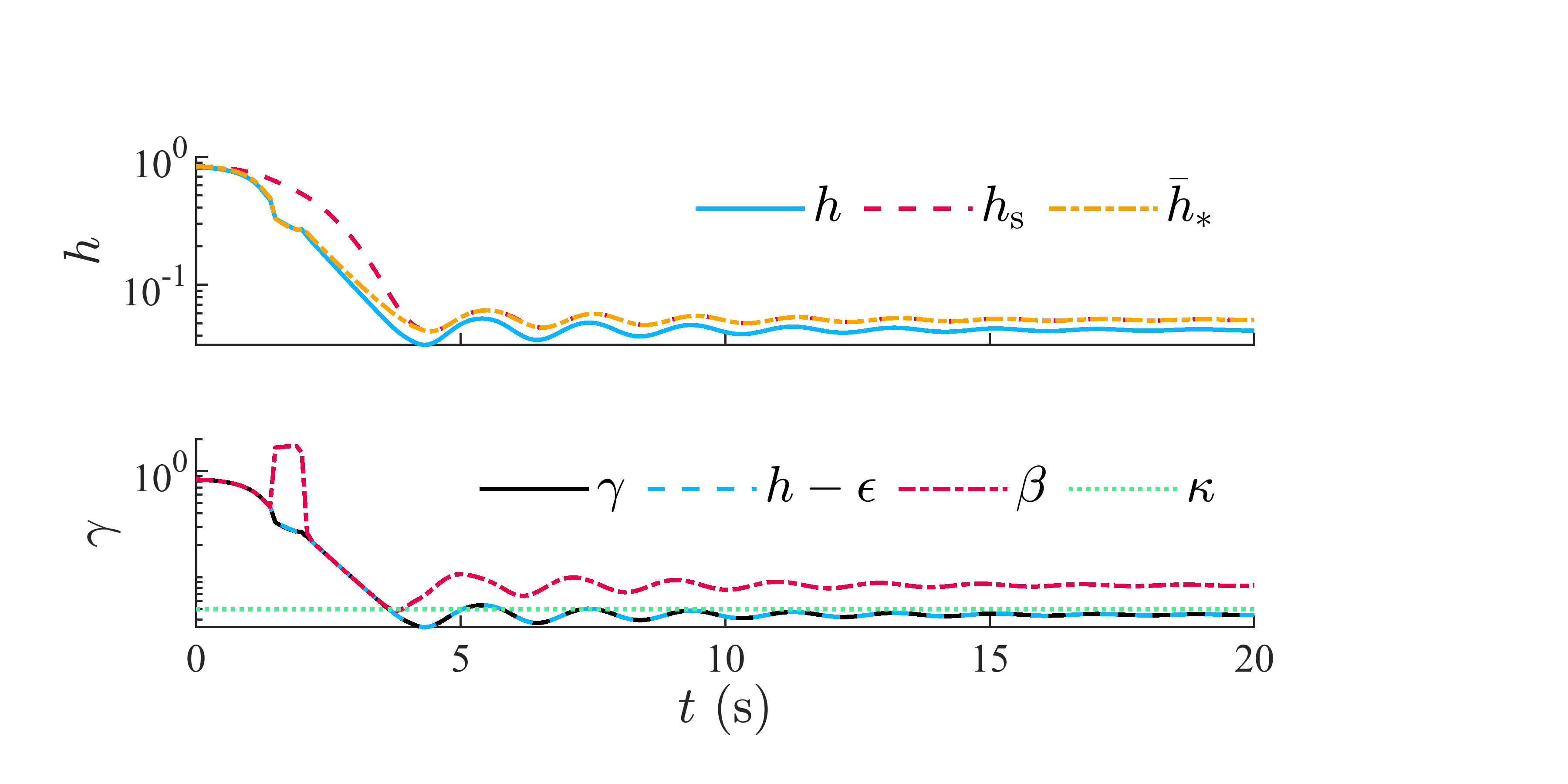}}
\caption{$h$, $h_\rms$, $\bar h_*$, $\gamma$, $h-\epsilon$, $\beta$ and $\kappa$ for $x_0=[0.5\,\,0]^\rmT$ and $\epsilon = 0$.}\label{fig:pendulum_extra}
\end{figure}

\Cref{fig:pendulum_states,fig:pendulum_extra} provide time histories for the case where $x_0=[\, 0.5\quad 0 \, ]^\rmT$ and $\epsilon = 0$.
\Cref{fig:pendulum_states} shows $\theta$, $\dot \theta$, $u$, $u_\rmd$, $u_\rmb$, and $u_*$. 
The top row of \Cref{fig:pendulum_extra} shows that $h$, $h_\rms$, and $\bar h_*$ are nonnegative for all time. 
The bottom row of \Cref{fig:pendulum_extra} shows $\gamma$, $h$, and $\beta$. 
Note that $\beta$ is positive for all time, which implies that~\eqref{eq:qp_softmin} is feasible at all points along the closed-loop trajectory. 
Since $\gamma$ is positive for all time but is less than $\kappa$ in steady state, it follows from  \eqref{eq:u_cases_softmin} that $u$ in steady state is a blend of $u_\rmb$ and $u_*$ (as shown in \Cref{fig:pendulum_states}). 
Note that $u_\rmb$ takes the pendulum back to $\SSS_\rmb$ but does not satisfy the objective of using instantaneously minimum control effort.

\begin{figure}[ht]
\center{\includegraphics[width=0.48\textwidth,clip=true,trim= 0in 0.28in 1.0in 1.0in] {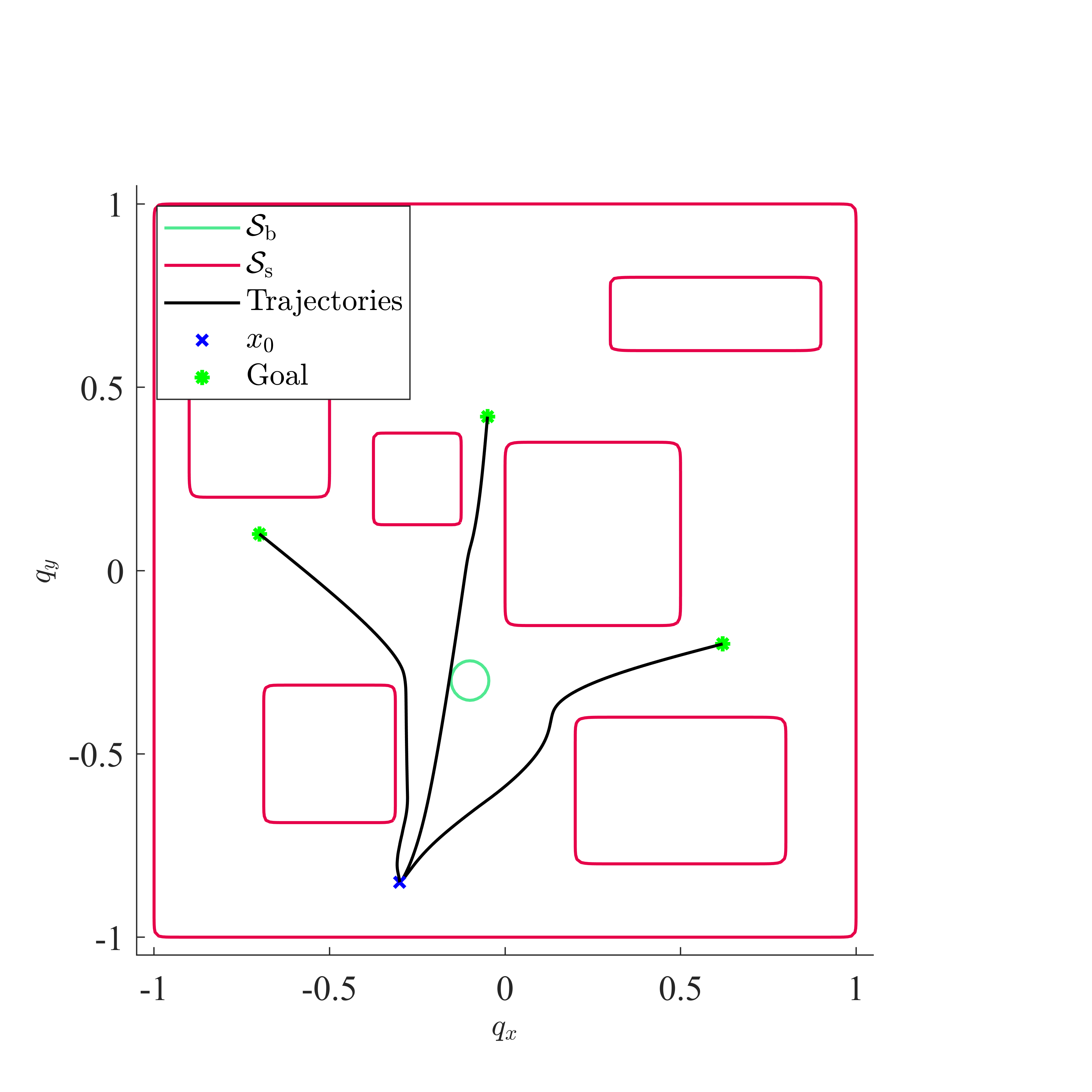}}
\caption{$\SSS_\rms$, $\SSS_\rmb$, and 2 closed-loop trajectories with Algorithms~\ref{alg:softmin}.}\label{fig:dp_map}
\end{figure}

\textbf{Ground Robot.} 
Consider the double-integrator ground robot modeled by \eqref{eq:dynamics}, where
\begin{equation*}
    f(x) = \begin{bmatrix}
    \dot q_x \\
    \dot q_y \\
    0 \\
    0
    \end{bmatrix}, 
    \quad
    g(x) = \begin{bmatrix}
    0 & 0\\
    0 & 0\\
    1 & 0 \\
    0 & 1
    \end{bmatrix}, 
    \quad
    x = \begin{bmatrix}
    q_x\\
    q_y\\
    \dot q_x\\
    \dot q_y
    \end{bmatrix}, 
    \quad
    u = \begin{bmatrix}
    u_1\\
    u_2
    \end{bmatrix}, 
\end{equation*}
and $q_x$ and $q_y$ are the positions in an orthogonal coordinate frame. 
Let $\bar u = 1$ and $\SU = \{ [ \, u_1 \quad u_2\, ]^\rmT \in\BBR^2: u_1,u_2 \in [-\bar u, \bar u] \}$. 
The safe set $\SSS_\rms$ projected into the $q_x$--$q_y$ plane is shown in \Cref{fig:dp_map}. 
Note that $\SSS_\rms$ is bounded in the $\dot q_x$ and $\dot q_y$ directions. 
The technical details of its construction are omitted for brevity. The backup control is 
\begin{equation*}
u_{\rmb}(x) = \begin{bmatrix}
\mbox{csat }K_1 (x - x_{\rmb})\\
\mbox{csat }K_2 (x - x_{\rmb})
\end{bmatrix},
\end{equation*}
where $\mbox{csat} \colon \BBR \to [-\bar u , \bar u ]$ is a continuously differentiable approximation of saturation, $K_1 = \matls -3.16 & 0 & -4.04 & 0 \matrs$, $K_2 = \matls 0 & -3.16 & 0 &  -4.04 \matrs$, and $x_{\rmb} = \matls -0.1 & -0.3 & 0 & 0 \matrs^\rmT$.

Let $h_{\rmb}(x) = c_\rmb - (x - x_{\rmb})^\rmT P_\rmb(x - x_{\rmb})$, where $c_\rmb = 0.0034$ and $P_\rmb \in \BBR^{4 \times 4}$ is determined using Lyapunov's direct method in order to ensure that Assumption~\ref{assump:S_b_contract} is satisfied. 
\Cref{fig:dp_map} shows as projection of $\SSS_{\rmb}$ into the $q_x$--$q_y$ plane.
The desired control is $u_\rmd = \matl{cc} K_1(x-x_\rmg) & K_2(x-x_\rmg) \matr^\rmT$, where $x_\rmg \in\BBR^4$ is the desired value of the state (i.e., the goal location).

We implement the soft-minimum BF control~\Cref{eq:h_softmin_def,eq:defS_softmin,eq:feas_check,eq:feasible_set_def,eq:gamma_def,eq:Gamma_def,eq:qp_softmin,eq:u_cases_softmin}, where $\rho = 100$, $\epsilon=0.1$, $\alpha =1$, and $\sigma$ is given by \Cref{ex:sigma} where $\kappa = 0.1$.
We let $\delta t= 0.02\,\rms$, $N = 30$ and $T_\rms = 0.1\,\rms$.

\Cref{fig:dp_map} shows the closed-loop trajectories for 3 different values of the goal $x_\rmg$.
In each case, $x$ converges to the goal $x_\rmg$ while satisfying safety and the actuator constraints.

\begin{figure}[ht]
\center{\includegraphics[width=0.48\textwidth,clip=true,trim= 0.24in 0.3in 0.65in 0.65in] {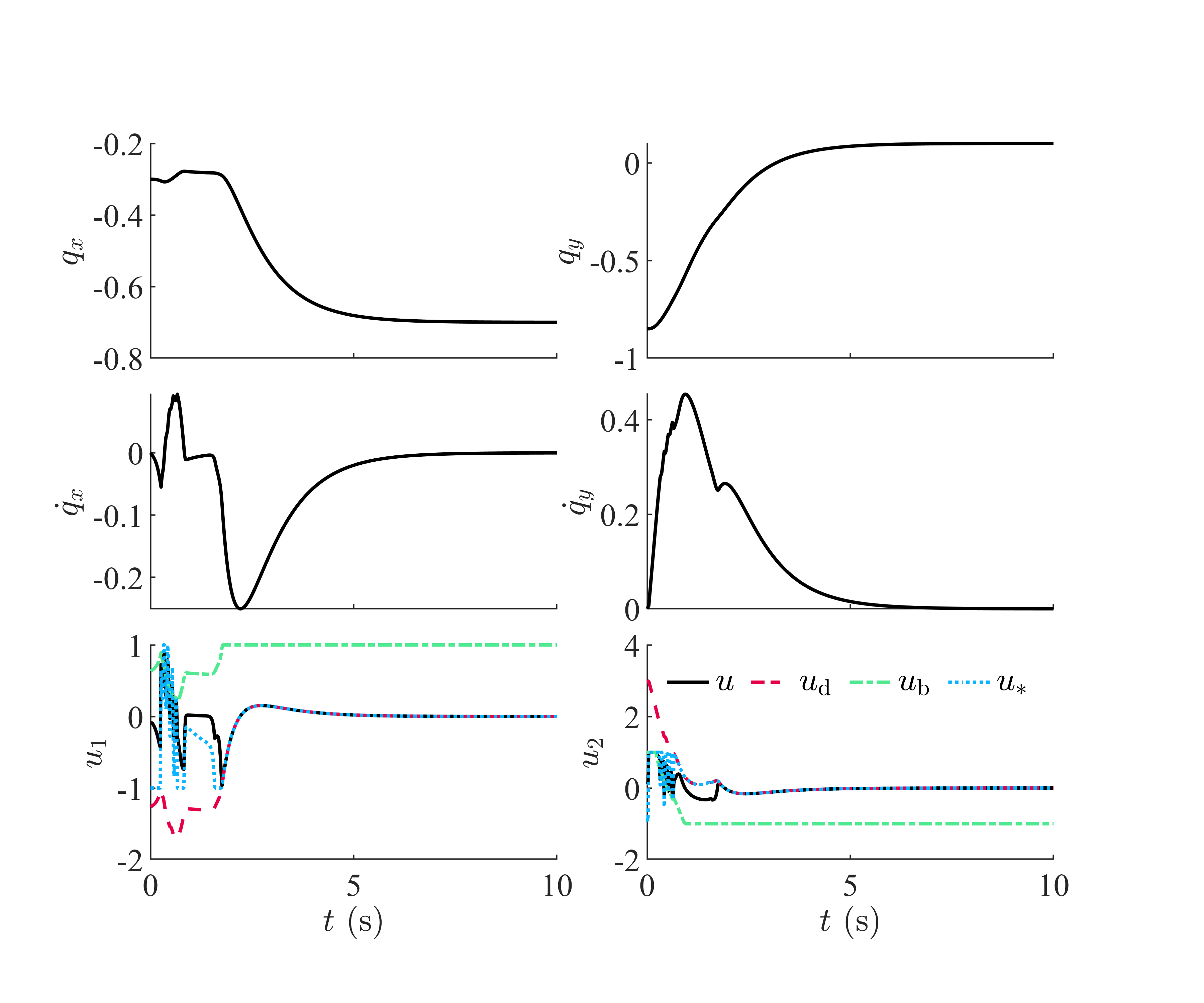}}
\caption{$q_x$, $q_y$, $\dot q_x$, $\dot q_y$, $u$, $u_\rmd$, $u_\rma$ and $u_*$ for Algorithm~\ref{alg:softmin} with $x_{\rmg}=[\,-0.7\,\,0.1\,\,0\,\,0\,]^\rmT$.}\label{fig:dp_states}
\end{figure}

\begin{figure}[ht]
\center{\includegraphics[width=0.48\textwidth,clip=true,trim= 0.3in 0.2in 1in 0.55in] {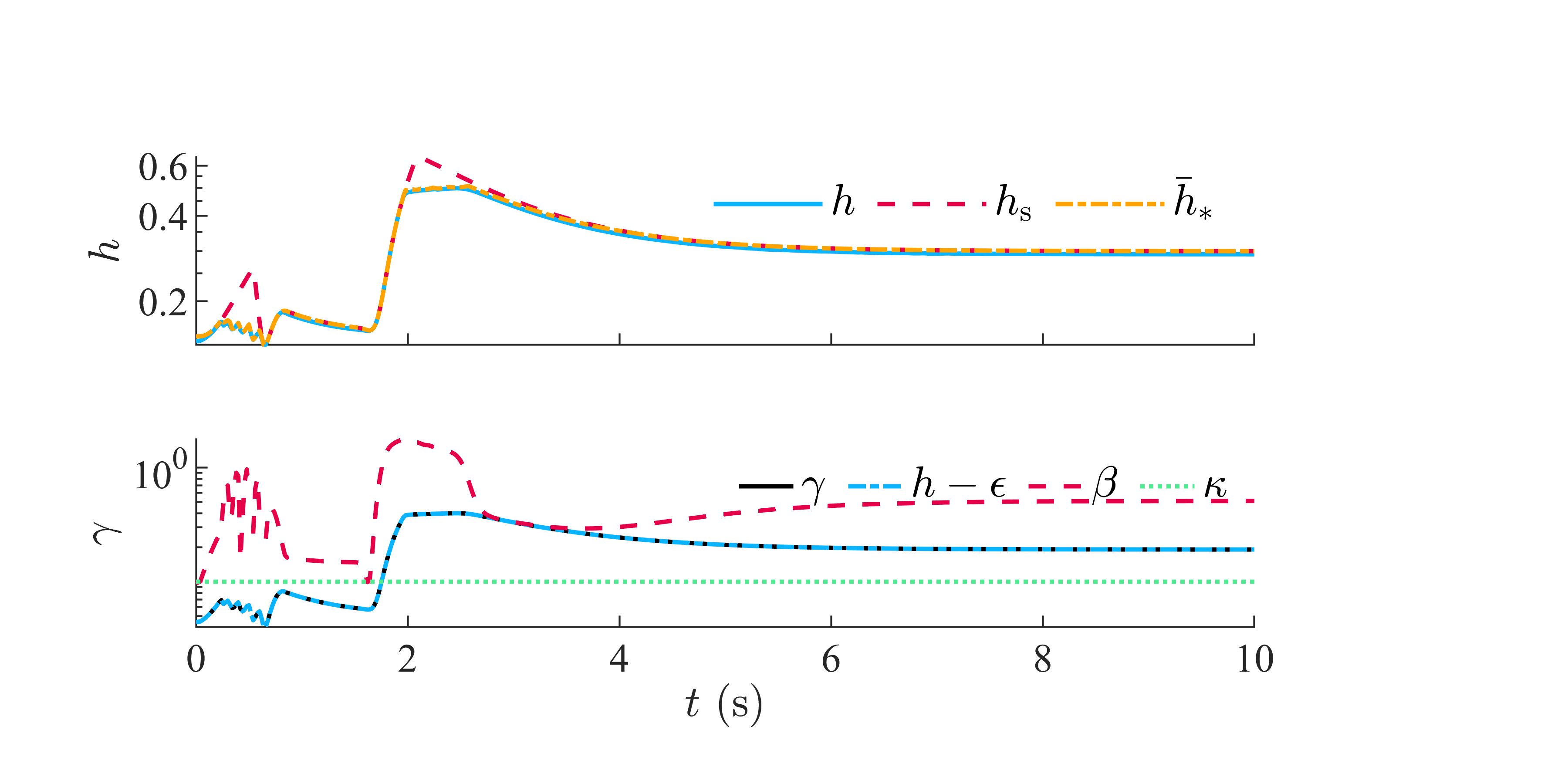}}
\caption{$h$, $h_\rms$, $\bar h_*$, $\epsilon$, $\gamma$, $h-\epsilon$, $\beta$ and $\kappa$ for Algorithm~\ref{alg:softmin} with $x_{\rmg}=[\,-0.7\,\,0.1\,\,0\,\,0\,]^\rmT$.}\label{fig:dp_extra}
\end{figure}

\Cref{fig:dp_states,fig:dp_extra} provide time histories for the case where $x_{\rmg}=\matls -0.7 & 0.1 & 0 & 0\matrs^\rmT$.
\Cref{fig:dp_states} shows $q_x$, $q_y$, $\dot q_x$, $\dot q_y$, $u$, $u_\rmd$, $u_\rma$, and $u_*$.
\Cref{fig:dp_extra} shows $h$, $h_\rms$, $\bar h_*$, $\gamma$, $h - \epsilon$, and $\beta$. 
For all $t \in [0,1.75]$, $\gamma < \kappa$ because $h - \epsilon < \kappa$.
Thus, during this time $u$ is computed from a blending of $u_\rmb$ and $u_*$ according to \eqref{eq:u_cases_softmin}.
For all $t > 1.75$, $\gamma > \kappa$, which implies that during this time $u = u_*$. 
In fact, $u_* = u_\rmd$ for all $t > 1.75$. 

\bibliographystyle{ieeetr}
\bibliography{backup_cbf.bib}

\end{document}